\documentclass[%
 reprint,prl,
 amsmath,amssymb,
 aps,superscriptaddress
]{revtex4-1}

\usepackage{graphicx}
\usepackage{dcolumn}
\usepackage{bm}
\usepackage{subfig}
\usepackage{float}

\usepackage{color}
\usepackage[normalem]{ulem}

\begin{document}

\preprint{APS/123-QED}

\title{Emergent Composite Particles from the Universal Exact Identities in Quantum Many-Body Systems with Generic Bilinear Interactions}

\author{Hui Li}
 \email{physicslihui@zju.edu.cn}
\affiliation{%
Institute for Advanced Study in Physics and School of Physics, Zhejiang University, Hangzhou 310027, China
}

\date{\today}

\begin{abstract}
A fundamental challenge in quantum many-body physics is to understand the universal properties of strongly correlated systems. In this work, we establish a universal and exact identity from the Dyson-Schwinger equations within the Keldysh-Schwinger field theory for systems with generic bilinear interactions. Our derivation demonstrates the emergence of composite particles as "elementary excitations", whose Green's functions definitively determine the original single-particle Green's function. This universal relation uniquely identifies the composite particles governing correlations and rigorously connects their spectra to the observable single-particle spectrum. Thus, our exact identity reveals a new pathway toward a paradigm for understanding many-body correlated systems.
\end{abstract}

\maketitle


\textit{Introduction}---
The accurate calculation of both single-particle and collective properties in quantum many-body systems remains a central challenge in condensed matter physics. For weakly correlated systems, Landau's Fermi liquid theory provides a remarkably successful paradigm, describing the low-energy physics in terms of long-lived quasiparticles that adiabatically connect to non-interacting electrons\cite{landau1959theory,coleman2015introduction}. This quasiparticle picture not only explains thermodynamic properties but is the foundation for understanding of transport and spectroscopic phenomena in diverse materials.

In stark contrast, strongly correlated systems, including cuprates\cite{keimer2015quantum,WXG2006_RMP}, heavy-fermion compounds\cite{kirchner2020colloquium,stockert2011unconventional}, and fundamental models like the Hubbard model\cite{qin2022hubbard}, lack such a universal description. Their complex phase diagrams, featuring unconventional superconductivity, pseudogap, strange metal behavior, and other strongly correlated phenomena, cannot be described by the quasiparticle paradigm. This complexity has driven the development of various many-body computational approaches, several representative categories of which include: (i) numerically exact methods, including density matrix renormalization group (DMRG)\cite{white1992dmrg,white1993dmrg,schollwock2011dmrg} and its higher-dimensional tensor network extensions\cite{banuls2023tensor,orus2019tensor,liu2017PEPS}, as well as various quantum Monte Carlo (QMC) approaches\cite{blankenbecler1981dqmc,scalapino1981dqmc,santos2003introductiondqmc,mishchenko2000diagrammatic,gull2011continuous}, (ii) dynamical mean-field theory (DMFT) and its cluster extensions\cite{rohringer2018diagrammatic,kotliar2001cellular,maier2005quantum,toschi2007dynamical}, and (iii) diagrammatic field-theoretic methods\cite{kohn1960Luttinger,baym1961conservation} such as the GW approximation\cite{Hedin,GW_review_2019,cGW_2023}, fluctuation-exchange approximation (FLEX)\cite{bickers1989FLEX}, two-particle self-consistent approach (TPSC)\cite{vilk1997TPSC}, and slave-particle formulations\cite{kotliar1986slave}. While each method has achieved significant success for certain models and parameter regimes, all face intrinsic limitations: DMRG is most effective in low dimensions; QMC suffers from the fermionic sign problem; DMFT struggles to capture non-local spatial correlations; and perturbative diagrammatic methods break down in the strong-coupling regime. As a result, a comprehensive and unified theoretical description for correlated systems remains elusive.

An important insight toward such a problem emerged from the work of Xiao-yong Feng\cite{feng2024inverted}, who discovered an "inverted duality" in the normal state of the Hubbard model—a rigorous relation between the electron and doublon Green's functions. This suggests that exact, universal identities may exist more generally in correlated systems. Motivated by this insight, we investigate a broad class of systems with generic bilinear interactions. Using the Keldysh-Schwinger field theory\cite{chou1985equilibrium,rammer2011quantum,stefanucci2013nonequilibrium,altland_simons_2010} and the Dyson-Schwinger formalism, we derive a universal and exact identity for the single-particle Green's function.

This identity provides a new perspective on the structure of quantum correlations. Most significantly, it naturally reveals the emergence of composite particles as "elementary excitation", whose Green's functions are shown to completely determine the original single-particle propagator. We demonstrate the power of this identity through a detailed analysis of the equilibrium finite-temperature case and its application to two distinct types of correlated systems. Crucially, this composite-particle paradigm is not restricted to fermionic or equilibrium settings; it holds universally for bosons, non-equilibrium scenarios, different phases, and systems with non-local interactions. Our work elucidates a universal structure underlying correlation effects, thereby offering a new foundational insight and a potential starting point for a future paradigm for strongly correlated quantum matter. 

\textit{General formalism}---
We start with a general Hamiltonian:
\begin{equation}
    \hat H=\sum_{ab}\hat c_{a}^\dagger t^{ab}\hat c_b+\frac{1}{2}\sum_{\alpha\beta}\hat\phi^{\alpha}V_{\alpha\beta}\hat\phi^\beta. \label{eq:Hamiltonian}
\end{equation}
Here, the label $a=(\vec x_a, i_a)$ is the degrees of freedom of the original particle $c$, containing both the space coordinate $\vec x_a$ and the flavor index $i_a$, such as the spin index $i_a=\sigma_a=\uparrow,\downarrow$.  $\hat\phi^\alpha=\sum_{ab}\hat c_{a}^\dagger\lambda^{\alpha ab}\hat c_{b}$ represents a bilinear interaction mode formed by the coupling of original particles via the structure tensor $\lambda$. Such operators resemble two-particle states or bosonic excitations, such as magnons, charge density. The Greek letter $\alpha$ labels the freedoms of the interaction mode, which may contain the space coordinate $\vec{x}_\alpha$ and the flavor for $\phi$. The tensor $\lambda^{\alpha ab}$ describes the coupling structure between the original particles and the interaction modes. For instance, in the case of spin interactions, $\phi^\alpha={ S}^{i_\alpha}(\vec x_\alpha)=\sum_{\sigma_a\sigma_b}\hat c^\dagger_{\sigma_a}(\vec x_\alpha)\tau^{i_\alpha}_{\sigma_a\sigma_b}\hat c_{\sigma_b}(\vec x_\alpha)$, the flavor index $i_\alpha = x, y, z$, the tensor $\lambda^{\alpha ab}=(\tau^{i_\alpha})_{i_a i_b}$ corresponds to the Pauli matrices. This Hamiltonian can describe both lattice and continuous systems, single-band or multi-band, local and long-range interactions, as well as bosonic and fermionic systems. It captures a wide range of condensed matter systems, such as but not limited to those with Hubbard, Coulomb, Yukawa, or spin interactions, in both fermionic and bosonic representations.

\begin{figure}
\includegraphics[width=0.45\textwidth]{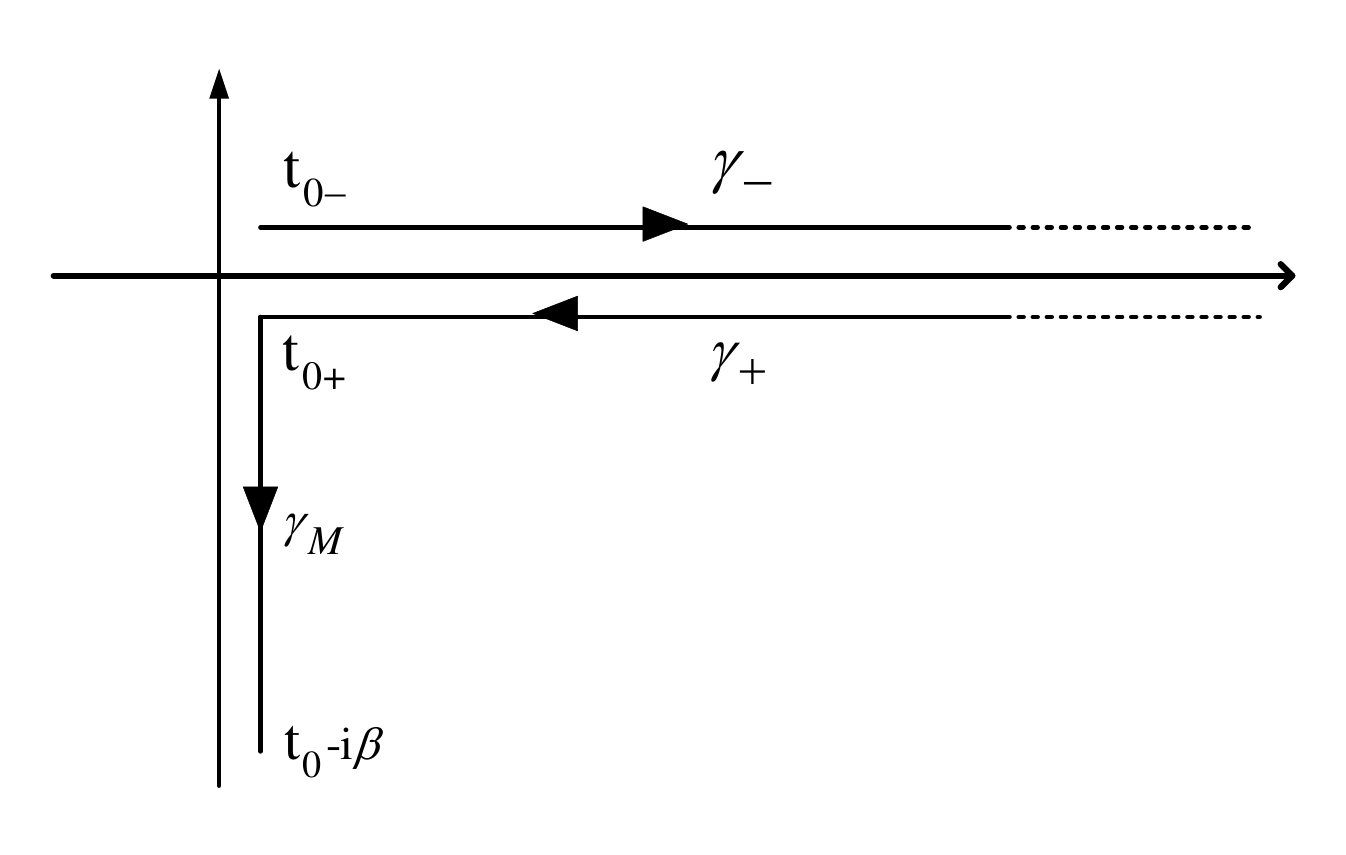}
\caption{The full Keldysh contour for finite-temperature quantum dynamics.\label{keldysh} }
\end{figure}

To derive the relations of the correlation function generally, we write Eq.~(\ref{eq:Hamiltonian}) in the Keldysh–Schwinger field formalism, with the action taking the form:
\begin{align}
S=&-\frac{1}{2}\psi_{A}(G_{0}^{-1})^{AB}\psi_{B}+\frac{1}{2}\phi^{\alpha}V_{\alpha\beta}\phi^\beta.
\end{align}
For fermions, the field $\psi$ is a Grassmann variable, and for bosons, it is a complex number. Exchange of field operators introduces a factor of $\xi = +1$ for bosons and $\xi = -1$ for fermions. Without loss of generality, we employ the general Nambu representation where the composite index $A = (C_A, a)$ consists of: a charge index $C_A$ (taking values $\psi^*$ and $\psi$ corresponding to creation and annihilation operator components, respectively), and a space-time-flavor index $a = (\vec{x}_a, t_a, i_a)$ that encompasses spatial coordinates, time, and flavor degrees of freedom. Similarly, the interaction mode index $\alpha$ also includes a time coordinate defined along the Keldysh contour\cite{stefanucci2013nonequilibrium}. The time variable $t_a$ is defined on the Keldysh contour $\gamma = \gamma_- \oplus \gamma_+ \oplus \gamma_M$, which consists of a forward branch $\gamma_-$ (from $t_{0-}$ to $\infty$), a backward branch $\gamma_+$ (from $\infty$ to $t_{0+}$), and an imaginary-time branch $\gamma_M$ (from $t_0$ to $t_0 - \mathrm{i}\beta)$, as shown in figure.~\ref{keldysh}. In this representation, the fully symmetrized mode operator is defined as $\phi^\alpha = \frac{1}{2} \psi_A \lambda^{\alpha AB} \psi_B$, where the tensor $\lambda^{\alpha AB}=\xi\lambda^{\alpha BA}$ ensures proper symmetrization.
We employ the Einstein summation convention throughout this work. The free part of the Keldysh action is given by
\begin{align}
S_0 =& \int_{t_0}^\infty dt\sum_{ab}[\psi^*_{a-}(t)(\mathrm{i}\partial_t\delta_{ab}-t_{ab})\psi_{b-}(t)\\
&- \psi^*_{a+}(t)(\mathrm{i}\partial_t\delta_{ab}-t_{ab})\psi_{b+}(t)] \nonumber\\
    &+ \int_{t_0}^{t_0-\mathrm{i}\beta} dt\sum_{ab}\psi^*_{a,M}(t)\left(\mathrm{i}\partial_t\delta_{ab} - t^{ab}-\mu\right)\psi_{b,M}(t),\nonumber
\end{align}
where $\mu$ is the chemical potential, and $\psi_{-},\psi_{+},\psi_M$ are the fields defined on different Keldysh branches $\{\gamma_- ,\gamma_+ ,\gamma_M\}$. The generating functional is
\begin{equation}
    Z=\int D\psi e^{\mathrm{i}S[\psi]}.
\end{equation}
The Green's function is given by $G_{AB} = \mathrm{i} \left\langle \psi_B \psi_A \right\rangle$, where the expectation value $\left\langle \cdots \right\rangle$ is defined as
\begin{equation}
  \left< \cdots \right>=\frac{1}{Z}\int D\psi \cdots e^{\mathrm{i}S}  
\end{equation}
The generalized Dyson–Schwinger (DS) equation is derived from
\begin{align}
\int D\psi \frac{\delta}{\delta\psi_A}[\mathcal F e^{\mathrm{i}S}]=0,
\end{align}
where $D\psi$ defines the measure of the functional integral, and $\mathcal F$ is an arbitrary functional. To obtain the one-body Green's function, we take the functional $\mathcal F=\psi_B$, which leads to
\begin{align}
    \delta^{A}_B+\mathrm{i}\xi\left< \psi_{B}\frac{\delta S}{\delta\psi_A} \right>=0.
\end{align}
This equation can be expanded as:
\begin{align}
    \delta^A_B-\mathrm{i}(G_{0}^{-1})^{AD}\left< \psi_{B}\psi_D \right>+\mathrm{i}\left< \psi_{B}\lambda^{\alpha AD}\psi_D V_{\alpha\beta}\phi^\beta \right>=0.\label{eq:DS1-final}
\end{align}
Then, we derive the second DS equation by choosing the functional $\mathcal F=\lambda^{\alpha AD}\psi_D V_{\alpha\beta}\phi^\beta$, which yields:
\begin{align}
&\lambda^{\alpha AD}\delta^E_DV_{\alpha\beta}\left< \phi^\beta\right>+\xi\lambda^{\alpha AD}\left< \psi_DV_{\alpha\beta}\lambda^{\beta FK}\psi_K\right>\delta_F^E\nonumber\\
&+\mathrm{i}\xi\left< \lambda^{\alpha AD}\psi_DV_{\alpha\beta}\phi^\beta\frac{\delta S}{\delta \psi_E} \right>=0\label{eq:DS2-original}
\end{align}
The last term in Eq.~(\ref{eq:DS2-original}) expands to:
\begin{align}
    &(G_0^{-1})^{FE}\left< \lambda^{\alpha AD}\psi_DV_{\alpha\beta}\phi^\beta\psi_F \right>\\
    &+\left< \lambda^{\alpha AD}\psi_DV_{\alpha\beta}\phi^\beta \lambda^{\gamma EF}\psi_F V_{\gamma\theta}\phi^\theta\right>,\nonumber
\end{align}
where a composite particle structure emerges. This motivates the definition of a composite particle operator:
\begin{equation}
    d^A= \lambda^{\alpha AD}\psi_DV_{\alpha\beta}\phi^\beta.
\end{equation}
This operator represents a bound state between a bare particle $\psi$ and the dynamical interaction mode $\phi$, forming the dominant "quasiparticle". Its emergence signifies the dressing of bare particles by interaction, making $d^A$ the fundamental, screened excitation that governs the system's correlation effects.

The corresponding "single-particle" Green's function of the composite particle is defined as
\begin{equation}
    \mathcal{C}^{AB}=\mathrm{i}\left< d^B d^A \right>.
\end{equation}
Consequently, the second DS equation can be expressed as:
\begin{align}
    &\lambda^{\alpha AE}V_{\alpha\beta}\left< \phi^\beta\right>-\mathrm{i}\xi\lambda^{\alpha AD}V_{\alpha\beta}\lambda^{\beta EK}G_{KD}\nonumber\\
    &+\mathrm{i}\xi(G_0^{-1})^{FE}\left< \lambda^{\alpha AD}\psi_DV_{\alpha\beta}\phi^\beta\psi_F \right>+\mathcal{C}^{AE}=0.\label{eq:DS2-final}
\end{align}
By comparing the Eq.~(\ref{eq:DS1-final},\ref{eq:DS2-final}), the term $\left< \psi_{B}\lambda^{\alpha AD}\psi_DV_{\alpha\beta}\phi^\beta\right>$ can be canceled, leading to a key identity:
\begin{align}
    &\lambda^{\alpha AB}V_{\alpha\beta}\left< \phi^\beta\right>-\mathrm{i}\xi\lambda^{\alpha AD}V_{\alpha\beta}\lambda^{\beta BK}G_{KD}+(G_0^{-1})^{AB}\nonumber\\
    &-\xi(G_0^{-1})^{AC}G_{CF}(G_0^{-1})^{FB}+\mathcal{C}^{AB}=0.
\end{align}
The first term, involving the expectation value of the interaction mode $\phi$, corresponds to the Hartree-type contribution. The second term represents the Fock-type contribution. The final term is the composite particle correlator $\mathcal{C}$, which emerges spontaneously in correlated systems. This identity establishes a relation between the one-body Green's function of the "bare" particle $\psi$ and the Green's function of the composite particle $d$. The composite particle is central to the correlated physics of the many-body system under both equilibrium and non-equilibrium.  And this identity can also be written as:
\begin{align}
    (G^{-1})^{AB}=\xi(G_0^{-1})^{AB}-\Sigma^{AB}_{\mathrm{H}}-\Sigma^{AB}_{\mathrm{F}}-\Sigma^{AB}_{\mathrm{C}},
\end{align}
where the self-energy is 
\begin{align}
    \Sigma^{AB}_{\mathrm{H}}=&\lambda^{\alpha AE}V_{\alpha\beta}\left< \phi^\beta\right>(G_0)_{ED}(G^{-1})^{DB},\nonumber\\
    \Sigma^{AB}_{\mathrm{F}}=&-\mathrm{i}\xi\lambda^{\alpha AD}V_{\alpha\beta}\lambda^{\beta EF}G_{FD}(G_0)_{EK}(G^{-1})^{KB},\nonumber\\
    \Sigma^{AB}_{\mathrm{C}}=&-\xi\mathcal{C}^{AD}(G_0)_{DF}(G^{-1})^{FB}.
\end{align}
It should be noted that the standard mean-field theory corresponds to the factorization approximation
\begin{align}
    \mathcal{C}^{AB}\approx \left< \lambda^{\gamma BE}V_{\gamma\theta}\phi^\theta \right>\left< \lambda^{\alpha AD}V_{\alpha\beta}\phi^\beta \right>G_{DE},
\end{align}
whereas fluctuations beyond this factorization give rise to higher-order correlation effects..

\textit{Identity for Equilibrium Systems at Finite Temperature with Translational Invariance}---
In most condensed matter systems, calculations are performed under the assumption of equilibrium at finite temperature and in the absence of $U(1)$ symmetry breaking. Therefore, we present the composite particle identity within the framework of the standard Matsubara formalism,
\begin{align}
S=&-\psi_a^*(G_{0}^{-1})^{ab}\psi_{b}+\frac{1}{2}\phi^\alpha V_{\alpha\beta}\phi^\beta,\label{eq:normal_action}
\end{align}
with the interaction mode $\phi^\alpha =\psi_a \lambda^{\alpha ab} \psi_b$. Here, the free propagator $G_0^{-1}$ can be written as
\begin{align}
(G_0^{-1})_{ab}=&(-\partial_\tau+\mu)\delta_{ab}-t_{ab}.
\end{align}
The partition function of this system is
\begin{equation}
    Z=\int D[\psi^*\psi] e^{-S[\psi^*,\psi]}.
\end{equation}
Following a derivation analogous to that in the Keldysh formalism, we obtain the identity within the Matsubara framework:
\begin{align}
\mathcal C^{ab}=&\lambda^{\alpha ab}V_{\alpha\beta}\left< \phi^\beta\right>+\lambda^{\alpha ad}V_{\alpha\beta}\lambda^{\beta cb}G_{cd}\nonumber\\
&+\xi(G_0^{-1})^{ab}+(G_0^{-1})^{ac}G_{cd}(G_0^{-1})^{db}.\label{eq:identity_M}
\end{align}
In the momentum-frequency space, where $k=(\vec k,\mathrm{i}\omega_k)$ denotes a combined momentum and Matsubara frequency variable, we employ the following Fourier transform conventions:
\begin{align}
    G(\vec x_a-\vec x_b;\tau_a-\tau_b)=&\sum_k e^{\mathrm{i}\vec{k}\cdot(\vec x_a-\vec x_b)-\mathrm{i}\omega_k(\tau_a-\tau_b)}G(k),\nonumber\\
    \lambda(\vec x_a\tau,\vec x_b\tau,\vec x_c\tau)=&\sum_{kq} e^{\mathrm{i}\vec{k}\cdot(\vec x_a-\vec x_b)-\mathrm{i}\omega_k(\tau_a-\tau_b)}\nonumber\\
    &e^{\mathrm{i}\vec{q}\cdot(\vec x_a-\vec x_c)-\mathrm{i}\omega_q(\tau_a-\tau_c)}\lambda(k,q).
\end{align}
Using these transformations, the identity Eq.~(\ref{eq:identity_M}) in the momentum-frequency space becomes:
\begin{align}
&\mathcal C^{i_ai_b}(k)=\lambda^{i_\alpha i_ai_b}(-k,k)V_{i_\alpha i_\beta}(p=0)\left<\phi^{i_\beta}(0)\right>\label{eq:identity-momentum}\\
&+\sum_q\lambda^{i_\alpha i_ai_d}(-k,-q)V_{i_\alpha i_\beta}(k+q)\lambda^{i_\beta i_ci_b}(q,k)G_{i_ci_d}(q)\nonumber\\
&+\xi(G_0^{-1})^{i_ai_b}(k)+(G_0^{-1})^{i_ai_c}(k)G_{i_ci_d}(k)(G_0^{-1})^{i_di_b}(k),\nonumber
\end{align}
where the summation means $\sum_q=\frac{1}{L^d\beta}\sum_{\vec q}\sum_{\mathrm{i}\omega_q}$ for the $d$ dimensional lattice system with size $L$.
Here, the Matsubara Green's functions are defined as:
\begin{equation}
    G_{ab}=\left< \psi_b^*\psi_a \right>,\:\mathcal{C}^{ab}=\left< d^{*b} d^a \right>.
\end{equation}
And the composite particle operators are:
\begin{equation}
    d^a= \lambda^{\alpha ad}\psi_dV_{\alpha\beta}\phi^\beta ,(d^a)^*=\lambda^{\alpha ba}\psi_b^* V_{\alpha\beta}\phi^\beta.
\end{equation}
Eq.~(\ref{eq:identity-momentum}) enables us to establish an exact relation between the spectral functions of the composite particle and the original particle. Under the physically reasonable assumption that the structure tensor $\lambda$, the interaction tensor $V$, the expectation value of the interaction mode $\langle \phi \rangle$, and the inverse free propagator $G_0^{-1}$ are all real-valued—which holds for most common systems—we derive the following relation:
\begin{align}
&\mathcal A^{i_ai_b}_{\mathrm{composite}}(k)=(G_0^{-1})^{i_ai_c}(k) A_{i_ci_d}(k)(G_0^{-1})^{i_di_b}\label{eq:spectral}\\
&+\sum_q\lambda^{i_\alpha i_ai_d}(-k,-q)V_{i_\alpha i_\beta}(k+q)\lambda^{i_\beta i_ci_b}(q,k) A_{i_ci_d}(q).\nonumber
\end{align}
Here, we define the spectral function of the particle $\psi$ and the composite particle $d$ as:
\begin{align}
     A_{ab}=-\frac{1}{\pi}\mathrm{Im} G_{ab},\quad\mathcal A^{ab}_{\mathrm{composite}}=-\frac{1}{\pi}\mathrm{Im} {\mathcal C}_{ab}
\end{align}
Equation~(\ref{eq:identity_M}) establishes a fundamental relation between the original and composite particles at thermal equilibrium. Thus, this identity provides a powerful framework for unveiling the spectral properties and characterizing the excitation spectra of composite particles in strongly correlated systems.

\textit{The application in various many-body systems at finite temperature}---
We first consider the Hubbard model, the minimal model of strongly correlated electron systems that captures essential many-body phenomena, including Mott physics, pseudo-gap\cite{qin2022hubbard}, magnetic ordering, strange metal\cite{Huang_2019}, and serves as a foundational model for exploring the mechanism of high-temperature superconductivity. Its Hamiltonian is given by
\begin{equation}
    \hat H=-\sum_{\vec k\sigma}\varepsilon_{\vec k}\hat c_{\vec k\sigma}^\dagger \hat c_{\vec k\sigma}+U\sum_{i}\hat n_{i\uparrow}\hat n_{i\downarrow}-\mu\sum_{i\sigma}\hat n_{i\sigma},
\end{equation}
where $\varepsilon_{\vec k}$ is the dispersion relation, and $U$ is the strength of the on-site interaction.The Matsubara action has the form as Eq.~(\ref{eq:normal_action}), with the flavor index corresponding to spin $i_a=\sigma_a,i_\alpha=\sigma_a$. 
Here, the interaction tensor takes the form
\begin{equation}
    V_{\alpha\beta}=U\delta(\vec{x}_\alpha,\vec x_\beta)\delta(\tau_\alpha,\tau_\beta)\delta_{\sigma_\alpha,\bar\sigma_\alpha},
\end{equation}
where $\bar\sigma$ denotes the opposite spin of $\sigma$. And the structure tensor is $\lambda^{\alpha ab}=\delta^{\alpha a}\delta^{\alpha b}$. By applying the Eq.~(\ref{eq:identity-momentum}), one can obtain the exact identity for the composite particle in the Hubbard model:
\begin{equation}
    Un_{\bar\sigma}-(G_0^{-1})_\sigma(k)+(G_0^{-1})^2_\sigma(k)G_\sigma(k)-\mathcal{C}_{\sigma}(k)=0.
\end{equation}
The free propagator is:
\begin{equation}
    G_{0}^{-1}(\vec k,\mathrm{i}\omega_n)=\mathrm{i}\omega_n-\varepsilon_k+\mu.
\end{equation}
And the composite particle operator takes the form:
\begin{equation}
    (d^*)_{\sigma}(\vec x,\tau)\equiv U\psi^*_{\sigma}(\vec x,\tau)n_{\bar\sigma}(\vec x,\tau).
\end{equation}
This identity reveals that the effective composite excitation, which encodes the correlation effects in the Hubbard model, corresponds to the original electron $\psi_\sigma$ screened by the local density of opposite-spin electrons $n_{\bar\sigma}$ on the same lattice site. This result is consistent with the identity derived by Feng in Ref.~[X]. A key advantage of our formulation is its natural extensibility to symmetry-broken phases, offering a unified framework for exploring correlation effects across both normal and ordered states.

We next study a three-dimensional electron gas with Yukawa interaction\cite{pines1962yukawa,mahan2013many}, a fundamental model for understanding screened Coulomb interactions in condensed matter systems. The Hamiltonian is given by
 \begin{align}
         \hat H=&\sum_\sigma\int d^3x \hat c^\dagger_{\sigma}(\vec x)\left(  -\frac{1}{2m_e}\nabla^2-\mu\right)\hat c_{\sigma}(\vec x)\nonumber\\
         &+\frac{1}{2}\frac{g}{4\pi}\sum_{\sigma\sigma'}\int d^3x d^3x'\hat n_{\sigma}(\vec x)\frac{e^{-\mathcal{M} |\vec x-\vec x'|}}{|\vec x-\vec x'|}\hat n_{\sigma'}(\vec x'),
 \end{align}
where $g$ denotes the interaction strength and $\mathcal{M}$ parametrizes the inverse screening length governing the spatial decay of the Yukawa potential.
The Masubara action has the form as Eq.~(\ref{eq:normal_action}), with the interaction tensor:
\begin{align}
 V_{\alpha\beta}=\frac{g}{4\pi}\frac{e^{-\mathcal{M} |\vec x_\alpha-\vec x_\beta|}}{|\vec x_\alpha-\vec x_\beta|}\delta(\tau_\alpha-\tau_\beta).
\end{align}

The tensor $\lambda$ takes the form $\lambda^{\alpha ab}=\delta^{\alpha a}\delta^{\alpha b}$. Then the composite particle is
\begin{align}
    d^a\equiv\sum_{\sigma_\beta}\int d^3x_\beta\frac{ge^{-\mathcal{M} |\vec x_\alpha-\vec x_\beta|}}{4\pi|\vec x_a-\vec x_\beta|}\psi_{\sigma_a}(\vec x_a,\tau_a) n_{\sigma_\beta}(\vec x_\beta,\tau_a)
\end{align}
This composite operator represents a fermionic quasiparticle dressed by a non-local cloud of screened density fluctuations, reflecting the dynamic reorganization of the electron gas due to the Yukawa interaction.
The exact identity here takes the form:
\begin{align}
\mathcal C_{\sigma}(k)=&\frac{g}{\mathcal{M}^2}n+\sum_q\frac{g}{(k+q)^2+\mathcal{M}^2}G_{\sigma}(q)\nonumber\\
&-(G_0^{-1})(k)+G_{\sigma}(k)G_0^{-2}(k)
\end{align}
where the free Green's function takes the form
 \begin{equation}
     (G_0^{-1})(\vec k,\mathrm{i}\omega_n)=\mathrm{i}\omega_n-\frac{\vec k^2}{2m_e}+\mu
 \end{equation}
This identity explicitly connects the properties of the dressed composite excitation to the bare electron Green’s function. The second term represents a generalized, momentum-dependent exchange process mediated by the Yukawa potential. It captures the non-local correlation where an electron scatters off the screened interaction field generated by its own distribution in momentum space.

\textit{Conclusion}---
In this work, we have derived a universal and exact identity for quantum many-body systems with generic bilinear interactions by employing the Dyson-Schwinger formalism within the Keldysh-Schwinger functional field theory. This identity fundamentally reveals that composite particles emerge naturally as the elementary excitations responsible for encoding correlation effects. We have demonstrated the power and universality of this framework by applying it to diverse systems, including the finite-temperature Hubbard model and the electron gas with Yukawa interactions, uncovering a unified composite-particle structure underlying their distinct physical behaviors.

It is crucial to emphasize that while our identity provides an exact basis for identifying these composite particles and their relationship to the original particles, direct numerical calculations for specific strongly correlated systems still rely on further approximation. In particular, future applications will need to develop schemes for the composite Green's function, such as truncating their equations of motion. The validity of such approximations will ultimately determine the accuracy of predicting both the composite and the single-particle properties in specific models.

Furthermore, the exact relation between the composite-particle spectrum and the single-particle spectrum offers a direct pathway for analyzing excitation properties across various models. A key advantage of our approach is its natural applicability to non-equilibrium problems, opening new avenues for studying quantum dynamics in strongly correlated systems. Together, these aspects establish a rigorous analytical foundation and provide a new perspective for understanding quantum many-body correlated effects, thereby offering a valuable starting point for future studies of strongly interacting matter.

\begin{acknowledgments}
This work is also supported by MOST 2022YFA1402701. The authors are very grateful to Wen-Yuan Liu, Xiaoyong Feng, Sheng Yang, and Hai-Qing Lin for valuable discussions.
\end{acknowledgments}
\nocite{*}

\bibliography{apssamp}

\end{document}